\documentclass[sigconf]{acmart}

\usepackage{amsmath,amsfonts}
\usepackage{algorithmic}
\usepackage{textcomp}
\usepackage{svg}
\usepackage{xcolor}
\usepackage{graphicx}
\usepackage{graphics}
\usepackage{float}
\usepackage{cclicenses}
\usepackage{xspace}
\usepackage{subfigure}
\usepackage{makecell}
\usepackage{multirow} 
\usepackage{mathtools}
\usepackage[T1]{fontenc}
\usepackage[ruled,linesnumbered]{algorithm2e}
\usepackage{cleveref}

\begin{document}

% \title{QuRA: Reinforcement Learning-Based Routing for Quantum Networks}
\title{Adaptive Entanglement Generation for Quantum Routing}

\author{Tasdiqul Islam}
\affiliation{
  \institution{University of Texas at Arlington}
  \city{Arlington}
  \state{TX}
  \country{USA}
}
\email{txi7184@mavs.uta.edu}

\author{Md Arifuzzaman}
\affiliation{
  \institution{Missouri University of Science and Technology}
  \city{Rolla}
  \state{MO}
  \country{USA}
}
\email{marifuzzaman@mst.edu}

\author{Engin Arslan}
\affiliation{
  \institution{Meta}
  \city{Menlo Park}
  \state{CA}
  \country{USA}
}
\email{enginarslan@meta.com}

\begin{abstract}
Entanglement generation in long distance quantum networks is a difficult process due to resource limitations and probabilistic nature of entanglement swapping. To maximize success probability, existing quantum routing algorithms employ computationally expensive solutions (e.g., linear programming) to determine which links to entangle and use for end-to-end entanglement generation. Such optimization methods however cannot meet the delay requirements of real-world quantum networks, necessitating swift yet efficient real-time optimization models. In this paper, we propose reinforcement learning (RL)-based models to determine which links to entangle and proactively swap to meet connection requests. We show that the proposed  RL-based approach is $20\times$ times faster compared to linear programming. Moreover, we show that one can take advantage of longevity of entanglements to (i) cache entangled link for future use and (ii) to proactively swap entanglement on high demand path segments, thereby increasing the likelihood of request success rates. Through comprehensive simulations, we demonstrate that caching unused entanglements lead to $10-15\%$ improvement on the performance of the state-of-the-arts quantum routing algorithms. Complementing the caching proactive entanglement swapping leads to further enhances the request success rate by up to $52.55\%$.
\end{abstract}

% \begin{CCSXML}
% <ccs2012>
%    <concept>
%        <concept_id>10003033.10003039.10003040</concept_id>
%        <concept_desc>Networks~Network protocol design</concept_desc>
%        <concept_significance>500</concept_significance>
%    </concept>
%    <concept>
%        <concept_id>10002950.10003714</concept_id>
%        <concept_desc>Mathematics of computing~Mathematical optimization</concept_desc>
%        <concept_significance>300</concept_significance>
%    </concept>
% </ccs2012>
% \end{CCSXML}

% \ccsdesc[500]{Networks~Network protocol design}
% \ccsdesc[300]{Mathematics of computing~Mathematical optimization}


\maketitle

% Custom commands
\newcommand{\name}{AEG\xspace}
\newcommand{\Lmax}{\mathrm{L}_\mathrm{max}}

\section{Introduction}
Unlike classical networks, connection establishment (i.e., entanglement generation) in long distance quantum networks \cite{duan2001long,muralidharan2016optimallong} is a difficult process that involves several probabilistic operations such as producing entangled qubit pairs for each link and swapping the entanglement over multiple links. Quantum routing algorithms aim to improve the success probability of long distance quantum communication by identifying which links to use for given set of requests considering success probabilities as well as resource limitations.

\begin{figure}[t]
  \centering
  \includegraphics[keepaspectratio=true,angle=0,width=0.5\linewidth]{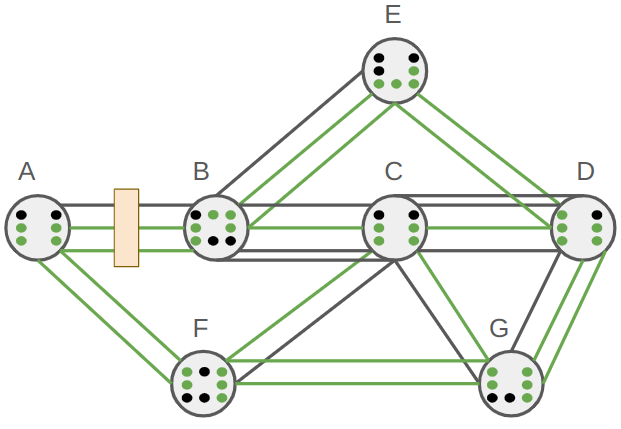}
  \vspace{-3mm}
  \caption{A simple quantum network with physical links with multiple paths between nodes and entangled photon source (EPS) in the middle of every link.}
  \label{fig:quantum_network}
  \vspace{-6mm}
\end{figure}

Figure~\ref{fig:quantum_network} depicts a simple quantum network with several quantum repeaters and Entangled Photon Sources (EPS). EPSes are positioned between nodes to produce entangled qubit pairs that are then transferred to adjacent nodes via quantum links. Once the link-level entanglement is established, swapping operation is performed to extend the entanglement to distant nodes. That is, entanglement between two adjacent link can be used to create entanglement between nodes that are two hop away. Then, the generated entanglement can be used to create entanglement between nodes that are three or more hops away. As the number of hops increases, the success probability decreases as the process can fail during any one of the entanglement generation or swapping operation.

Quantum routing algorithms play an essential role for quantum networks as they optimize which links to entangle and which routes to use for given connection requests, taking uncertainties introduced by quantum noise during the entanglement swapping process into account~\cite{yang2022online,shi2020qcastpass,zhao2021reps,zhao2022see,huang2022seer,chen2024redp,farahbakhsh2022opportunistic} . Typically, they follow a two-step process to establish an end-to-end entanglement as (i) selection of links for entanglement generation and (ii) selection of entangled links to establish end-to-end entanglement through entanglement swapping. Most quantum routing algorithms rely on compute-intensive optimization methods such as integer programming to optimize link and path selection decisions. For example, REPS algorithm~\cite{zhao2021reps} uses Integer Linear Programming both for links and path selection. However, such compute-intensive methods fail to produce results in a timely manner even for small networks with 30--40 nodes and 10--20 requests. As an example, it takes REPS algorithm around $140$ seconds to select which links to entangle and around $180$ to identify which set of links to extend the entanglement end-to-end. In addition, existing quantum routing algorithms fall short to achieve high success ratio due to not utilizing generated entanglements efficiently between consecutive time slots. Therefore, it is necessary to develop fast yet highly efficient quantum routing algorithms to pave the way for large scale quantum networks.

In this paper, we introduce \name to optimize the routing in quantum networks. \name proposes two improvements as follows: First, it introduces reinforcement learning based link selection strategy to quickly find which links to entangle. Unlike linear programming based solutions, reinforcement learning models can make prediction in the order of few seconds. Second, it innovates entanglement caching and proactive swapping methods to take advantage of unused entanglements for future requests. The rationale behind this approach is that since entanglement creation is a probabilistic process, caching unused entanglement as long as possible can increase the success rate. \name further takes advantage of unused entanglements to create multi-hop entanglements through entanglement swapping. This helps us to shorten the distance between end nodes, thereby improving the success rate of future connections.

Since \name does not yet offer a new solution for path selection algorithm, we integrated it into quantum routing algorithm REPS~\cite{zhao2021reps} such that it can use linear programming for path selection. Our evaluations show that the reinforcement learning based link selection of \name results in similar performance as REPS with 20 times faster in terms of execution time. We also show that applying our link selection approach with simply caching unused entangled qubits for several time slots (e.g., $5$--$10$ time slots), \name is able to outperform REPS by 20\% in terms of request satisfaction rate. Taking it a step further by proactive entanglement swapping across multiple commonly used links lets \name to attain $61\%$ improvement over REPS.

\section{Related Work}

Various routing algorithms have been suggested for quantum communication, including Q-CAST \cite{shi2020qcastpass}, REPS \cite{zhao2021reps}, SEER \cite{huang2022seer}, DQRA \cite{le2022entanglementdeepqrl}, and SEE \cite{zhao2022see}, all aimed at maximizing throughput and reducing idle time. Most of them follow the four phase architecture discussed earlier where in the second phase they choose links for entanglement generation and works in time-slot manner. In \cite{yang2022online} they select links for a request on its arrival as this works as online algorithm. In \cite{santos2023shortest,van2013pathqdijkstra} they use shortest path approach to find paths and generate entanglements for those paths.

\begin{figure}
\begin{center}
\subfigure[A non optimal Entanglement Generation]{
\includegraphics[keepaspectratio=true,angle=0,width=.5\linewidth] {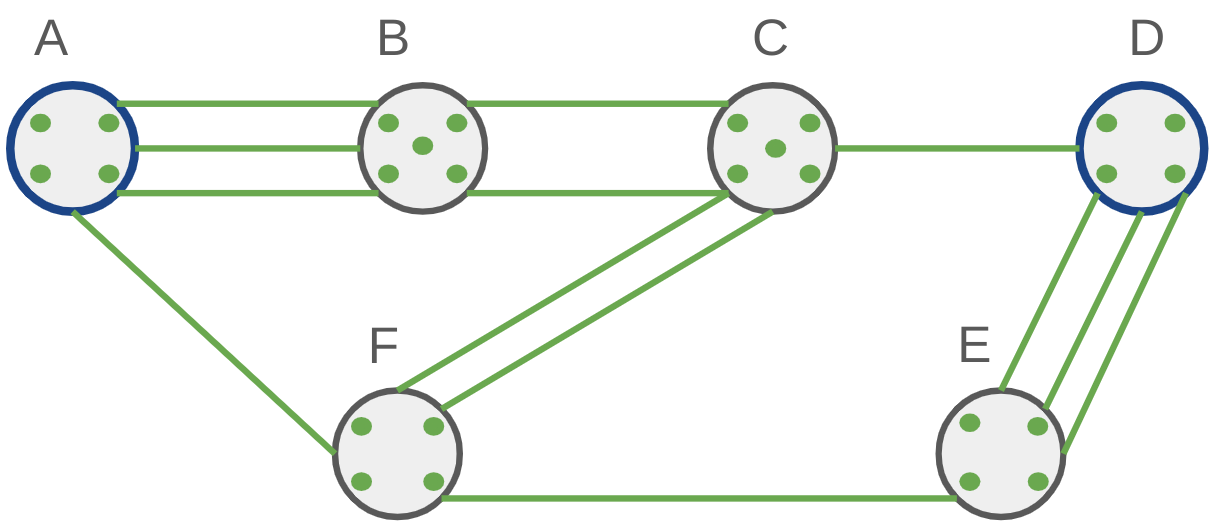}
\label{fig:mot1}}
\hspace{0mm}
\subfigure[An optimal Entanglement Generation]{
\includegraphics[keepaspectratio=true,angle=0,width=.5\linewidth] {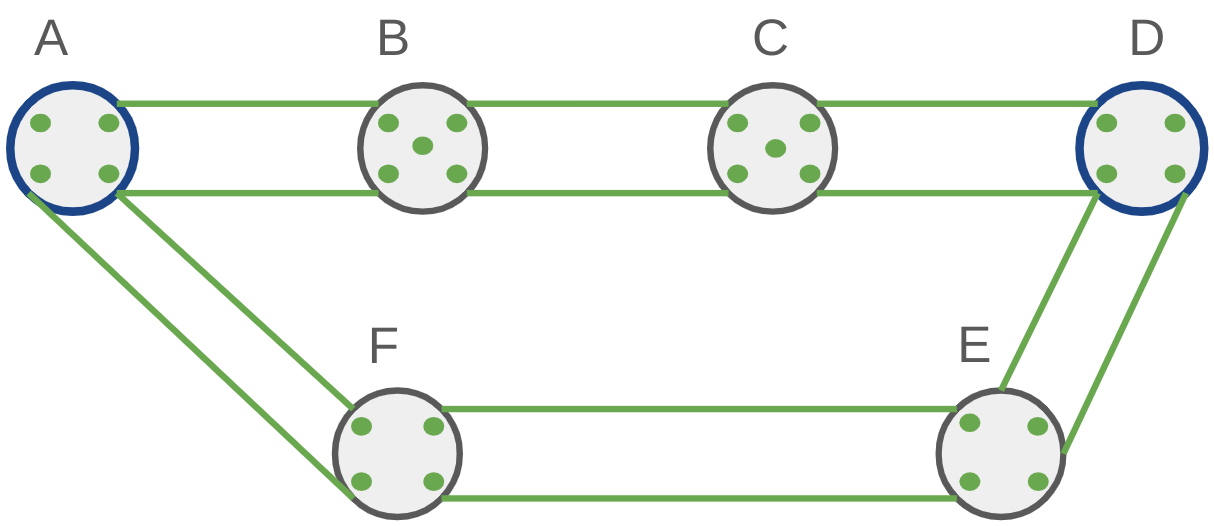}
\label{fig:mot2}}
\vspace{-2mm}
\caption{Optimal path selection for request A-D depends on which links are selected to generate entanglement}
\vspace{-6mm}
\label{fig:motivation}
\end{center}
\end{figure}

Choosing the right links for entanglement is important; to show this importance, Figure~\ref{fig:motivation} compares two cases for a single request between nodes A and D. In the first case (Figure~\ref{fig:mot1}), the selected links lead to a less reliable path. In the second case (Figure~~\ref{fig:mot2}), the chosen links form a more effective path. This example highlights how smart link selection can improve the overall performance of the network and motivates the need for better strategies for entanglement generation.

To select optimal links for generating entanglement REPS \cite{zhao2021reps} and SEE \cite{zhao2022see} use Integer Linear Programming where REPS \cite{zhao2021reps} tries to select links only between neighbor nodes but  SEE \cite{zhao2022see} considers all-optical switching to select non neighbor nodes and tries to generate entanglement between them. The constraints for Integer Linear Programming depends on the number of requests that causes more complexity with large number of requests.

To increase success probability SEER \cite{huang2022seer} considers dividing the main path into two by selecting an intermediate trusted node. The intermediate node is chosen based on social relationships within the network. To select connection between source to the intermediate node then to destination it considers the shortest path and from that shortest path links are selected for generating entanglement. To find the shortest paths it uses a greedy algorithm which executes faster than Integer Linear Programming. Based on the successful entangled links swapping happens to establish long entanglement which follows the shortest path also. Link selection and swapping both depend on the selected shortest paths.

% SEER \cite{huang2022seer} enhances source-destination entanglement reliability by dividing the main path into two. The qubit is initially sent to an intermediate trusted node, which then forwarded it to the final destination in the subsequent time slots. The intermediate node is chosen based on social relationships within the network. Dividing a long path into two segments reduces the path length, thereby increasing the success rate of entanglement establishment. On the other hand, it takes longer (i.e., at least two time slot) to complete a given request. SEER employs a greedy algorithm for path determination thus it executes faster than Linear Programming.

% DQRA \cite{le2022entanglementdeepqrl} employs a deep neural network to observe the current network states for scheduling requests, which are subsequently routed via a qubit-preserving shortest path algorithm.

Deep reinforcement learning was applied to schedule requests in DQRA \cite{le2022entanglementdeepqrl}, but for routing it uses a greedy shortest path algorithm. Links are selected for entanglement from the shortest path in this work. Another Deep reinforcement learning based approach has been proposed in \cite{roik2024routingrl}, but they do not consider multiple requests simultaneously.

Focusing on establishing entanglement between source-destination pairs, Q-CAST \cite{shi2020qcastpass} selects multiple redundant paths and selects links for entanglement from these paths. As this algorithm finds links for entanglement using a greedy algorithm, it also executes faster than Integer Linear Programming but does not perform similar or better than REPS~\cite{zhao2021reps} or SEE~\cite{zhao2022see}.

Since recent studies show that entanglement can lasts for several seconds~\cite{bartling2022entanglement1min,bradley2019entanglement10sec}, in \cite{proactiveswapping}, authors examined the impact of caching the unused entangled links on the performance of existing algorithms. With caching mechanism for entangled links they also showed the impact of proactive swapping which reduces the average distance of the networks resulting better performance for two existing routing algorithms. Deep Reinforcement Learning was applied to select optimal pairs of nodes for proactive swapping. Next, while REPS \cite{zhao2021reps} can produce near optimal performance using Integer Linear Programming (ILP), it is challenging to apply the solution for large number of requests in network with high number of nodes. In \name we replace ILP based link selection process with reinforcement learning, as it's batch prediction capabilities to select all potential links for entanglement at once make entire process exponentially faster and scalable for networks with high traffic.

\section{System Design of \name}

The existing quantum routing algorithms generally adopt a two-step process when handling connection requests. They are (i) selecting links to generate entanglement and (ii) finding path(s) using successfully entangled links to extend the entanglement from link level to end-to-end~\cite{shi2020qcastpass}. Since entanglement generation and swapping operations are probabilistic, quantum routing algorithms choose multiple links to entangle and swap for a given request since connection attempt on primary path may fail. Then, entangled qubit pairs are created for each link using Entanglement Photon Sources (EPS) \cite{zhao2021reps} whose success rate is contingent upon the inherent probability associated with each link. From the pool of successfully entangled links, an optimal path is selected for entanglement swapping. Similar to the entanglement generation, entanglement swapping operation can also fail due to the possibility of failure in the Bell State Measurement (BSM) \cite{kim2001quantumbsm} process. Ultimately, if at least one of the selected paths is successful in creating entanglement on all links and conducting BSMs, the request is deemed to be successful. 

Existing quantum routing algorithms discard unused entanglements between consecutive time slots, thus they perform entanglement generation and swapping operations in each round from scratch. However, previous studies showed that entanglement between two qubits can endure for up to ten seconds~\cite{bradley2019entanglement10sec}. Hence, we propose to cache unused entangled links for a few time slots (e.g., 10 time slots) to better utilize resources and improve the performance of routing algorithms~\cite{proactiveswapping}. Moreover, state-of-the-art quantum routing solutions rely on expensive computational solutions such as linear programming for optimal link and path selection, which are not scalable especially for large scale networks due to slow nature of these optimization algorithms. Hence, we introduce reinforcement-learning based link selection model to speed up the process.

\subsection{Reinforcement Learning-Based Link-Level Entanglement Generation}
Limited quantum memories and probabilistic nature of entanglement generation create a significant challenge for establishing end-to-end connections in quantum network. For a given set of requests and resources, Integer Linear Programming (ILP) can be applied to select which link to entangle and swap to meet as many requests as possible. While ILP yields high performance due to transparent mathematical optimization, it can take a long time to find a solution in large scale networks. Hence, it is not suited for problems that require real-time decisions. Reinforcement Learning (RL) based algorithms have been widely used to solve complex optimization problems. A trained RL model can directly predict all potential links quickly by avoiding the cost of repeated calculations. In particular, Deep Q-learning (DQRL) is known to attain high performance and scalability with the help of deep learning based predictions. However, RL models require a clear definition of states, actions and rewards function that accurately reflect the core aspects of the networking environment and routing objectives to achieve high performance. In our problem, we define them as follows:

%Deep Q-learning (DQRL) is known to scale well  based optimizer to select the optimal links for entanglement generation. 

\begin{figure}
\begin{center}
%\vspace{-2mm}
\includegraphics[keepaspectratio=true,angle=0,width=0.6\linewidth] {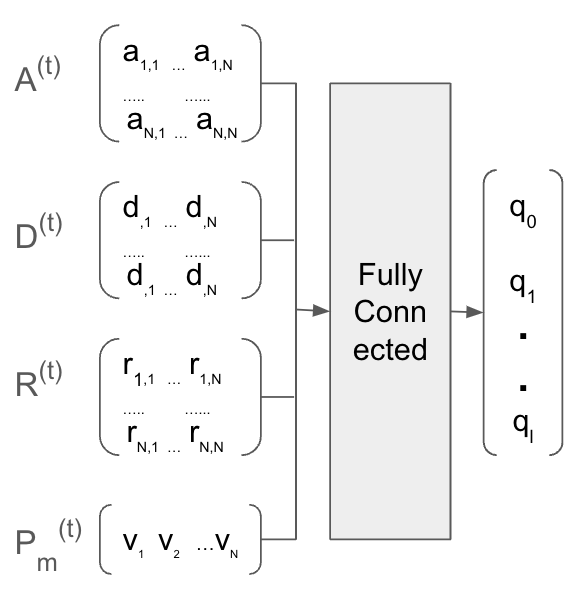}
\vspace{-3mm}
\caption{A sample input and output of the Deep Q-Learning. It takes network topology, distances among nodes, requests and entanglement edges as input and output a list of potential links for entanglement generation}
\label{fig:dqrl}
\vspace{-6mm}
\end{center}
\end{figure}

\begin{itemize}
    \item \textit{States:} The DQRL state space is consist of four components:
        \begin{enumerate}
            \item \textit{Topology:} The connectivity of the network $G^{(t)}(V,E^{(t})$ at time slot $t$ is calculated as a subset of the topology $G$. Only links that not entangled already are considered as edges. Then, we extract the adjacency matrix $A^{(t)} = (a_{i,j})_{1\leq i,j \leq |V|}$ of $G^{(t)}(V,E^{(t})$ where  $a_{i,j}$ is the number of links between nodes $i$ and $j$. This is a $N \times N$ matrix where $N$ is the number of nodes.
            \item \textit{Link Distance:} Distance of links impacts successful entanglement generation. Thus, we consider the distances between links in our state and represent them as adjacency matrix $D^{(t)} = (d_{i,j})_{1\leq i,j \leq |V|}$ where  $d_{i,j}$ is the distance between nodes $i$ and $j$.
            \item \textit{Requests:} The new requests at time slot $t$ along with the remaining requests from previous time slots are represented a $N \times N$ matrix which is denoted as $R^{(t)}=(r_{s,d})_{1\leq s,d \leq |V|}$ where $r_{s,d}$ is the number of requests between source and destination nodes. 
            \item \textit{Edges:} The $m^{th}$ pair $(n_1, n_2)$ from the set of edges in the topology that will be chosen to create entanglement is represented by a binary vector $P_m^{(t)}$, where each element $v_{i}$ is either 1 (if $i$ belongs to $(n_1, n_2)$) or 0 otherwise.
            
        \end{enumerate}
\end{itemize}
The above four components make up the input state for $m^{th}$ pair $S_m^{(t)}= [A^{(t)};D^{(t)};R^{(t)};P_m^{(t)}]$

\begin{itemize}
    \item \textit{Output Actions:} The Deep Q neural network outputs a Q-value vector for each link in each time slot $Q_m^{(t)} = [q_0^{(t)} , q_1^{(t)}....., q_l^{(t)} ]$ of length $l+1$ where $l$ is the maximum link capacity between two nodes and the action $a_m^{(t)} = argmax_{i \in {0 ,1}}(q_i^{(t)})$ is the number of links we'll try to generate entanglement for the $m^{th}$ pair or if $a_m^{(t)}$ is greater than maximum number of links between the pair then will try to entangle all the links. Figure \ref{fig:dqrl} shows an example of the input and output.
    \item \textit{Rewards:} We define rewards and penalty as follows.
    \begin{enumerate}
        \item We assign a positive reward ($R(s,a) > 0$) for an action (i.e., choosing a link to entangle) if a selected link is used for a request either directly or as part of link segments that is explained in Section~\ref{sec:pes}.
        \item We assign a negative reward (($R(s,a) < 0$) to an action if selected link expires (i.e., lifespan ends) without being used for a connection request. This penalty ensures that the model will try avoid choosing links that are not likely to be used in the future.
    \end{enumerate}
\end{itemize}

\textbf{Model Training:} Instead of using the traditional train-then-evaluate method,  we design the algorithm as an online optimizer with both exploration and exploitation phases \cite{ishii2002controlexplore}. Exploration phase helps the model to stay up-to-date with evolving networking dynamics. The exploitation phase, on the other hand, aims take advantage of current model weights to predicts optimal links to entangle in any specific time slot. In our implementation, we maintain two versions of Q network for model training, a prediction network and a target network. The Q-values are obtained from the prediction network which is trained using the Stochastic Gradient Descent algorithm \cite{bottou1991stochastic}. The target network's weights are updated periodically with those of the prediction network, typically at every few hundred time steps. The target network is exclusively utilized for forecasting future Q-values. While this process is expected to repeat indefinitely, we used a model with fixed number of iterations to compare results with the state-of-the-arts solutions.

\subsection{Proactive Entanglement Swapping}\label{sec:pes}
Existing routing algorithms discard the unused entanglements which necessitates new entanglement generation in each time slot. Taking advantage of longevity of entanglement, we aim to maximize the success of connection establishment by caching unused entangled links.  Since entanglements cannot be cached forever, we mark the entanglement generation time for each unused link to keep track of their remaining lifespan. If a cached entanglement is not utilized within its lifetime, it is destroyed and the associated memory is freed.

% Consider a scenario with four requests (A-F, A-E, B-E, B-F in Figure~\ref{fig:cache_0}) in a network where all links have three quantum entanglement capacity. Assume a quantum routing algorithm determined optimal routes as shortest path between source and destination. Figure~\ref{fig:cache_1} shows the successfully entangled links after the entanglement attempts. The routing algorithm then chooses paths based on successful entanglements extend the entanglement end-to-end via entanglement swapping. Due to capacity limitations and failure during entanglement swapping across links, assume A-E and B-F requests are successful and A-F and B-E requests are unsuccessful as depicted in Figure \ref{fig:cache_2}. Since two entanglements were created on links A-C and B-C while only one of them are used for A-E and B-F connections, the second ones can be cached for future requests. 

In addition to caching unused entanglements at the link-level,  we also conduct entanglement swapping across neighboring links to extend the entanglement from single link to multiple links (i.e., segments). These segments can then be used to meet future requests. Entanglement swapping over multiple links reduces the average path length, thereby increasing the probability of successful entanglement for future requests. Additionally, it creates additional routes between source and destination nodes, which increases the success probability of connection requests. On the other hand, the capacity of quantum repeaters are limited, thus, it is important to create entangled segments judiciously to avoid running out of quantum memory which could leave insufficient resources for future time slots. That is, since proactive entanglement swapping cannot guarantee a perfect prediction, it is important to reserve some links to meet ``unexpected'' requests in upcoming time slots. Therefore, we implemented another DQRL method to predict which segments to proactively establish entanglement such that the success rate of future requests can be improved. We define its state, actions, and rewards as follows:

\begin{itemize}
    % \item \textit{Environment:} Previous studies proposed both linear programming (REPS~\cite{zhao2021reps}) and heuristic models (SEER~\cite{huang2022seer}) for routing. We incorporated our proposed caching and proactive entanglement swapping approaches into these existing solutions. Hence, the routing algorithm (SEER or REPS) acts as the environment for the Q-learning model.
    \item \textit{Environment:} We incorporated our caching and proactive entanglement swapping approach into an existing quantum routing algorithm REPS~\cite{zhao2021reps}. REPS uses ILP to find optimal paths for a given set of connection requests. Hence, the routing algorithm REPS acts as the environment for the Q-learning model.
    \item \textit{States:} DQRL state space is consist of three components:
        \begin{enumerate}
            \item \textit{Topology:} The connectivity of the network $G^{(t)}(V,E^{(t})$ at time slot $t$ is calculated as a subset of the topology $G$. Only entangled links are considered as edges. Then, we extract the adjacency matrix $A^{(t)} = (a_{i,j})_{1\leq i,j \leq |V|}$ of $G^{(t)}(V,E^{(t})$ where  $a_{i,j}$ is the number of entangled links between nodes $i$ and $j$. This is a $N \times N$ matrix where $N$ is the number of nodes.
            \item \textit{Requests:} The new requests at time slot $t$ along with the remaining requests from previous time slots are represented a $N \times N$ matrix which is denoted as $R^{(t)}=(r_{s,d})_{1\leq s,d \leq |V|}$ where $r_{s,d}$ is the number of requests between source and destination nodes. 
            \item \textit{Segments:} The $m^{th}$ pair $(n_1, n_2)$ that will be chosen to create proactive entanglement is represented by a binary vector $P_m^{(t)}$, where each element $v_{i}$ is either 1 (if $i$ belongs to $(n_1, n_2)$) or 0 otherwise.
        \end{enumerate}
        The above three components make up the input state for $m^{th}$ pair $S_m^{(t)}= [G^{(t)};R^{(t)};P_m^{(t)}]$
    \item \textit{Output Actions:} The Deep Q neural network outputs a Q-value vector for each segment in each time slot $Q_m^{(t)} = [q_0^{(t)} , q_1^{(t)} ]$ of length 2 and the action $a_m^{(t)} = argmax_{i \in {0 ,1}}(q_i^{(t)})$ is applied on the $m^{th}$ pair. 
    % \item \textit{Training Algorithm:} Two versions of Q networks are maintained in the Deep Q-learning method for model training. They are a prediction network and a target network. The Q-values are obtained from the prediction network, which undergoes training using the Stochastic Gradient Descent Algorithm. The target network's weights are updated periodically with those of the prediction network, typically at every few hundred time steps. The target network is exclusively utilized for forecasting future Q-values.
\end{itemize}

\textit{Model Training:} The model training is similar to our previous exploration/exploitation model for link selection for entanglement generation. Deep Neural Network is evaluated for every segment to determine whether or not to swap the entanglement. Ideally, these decisions (i.e., actions) should be executed sequentially such that state changes caused by one action (e.g., reduction of available link capacity) can be reflected in the evaluation of another action. However, sequential execution introduces two challenges. First, evaluation order of actions will affect the performance since available capacity will decrease as more actions are selected, giving advantage to actions that are evaluated earlier. Second, it takes long time (up to 30 seconds) to evaluate DNN model sequentially for all possible actions. Hence, we evaluate actions in parallel, then choose the ones with the highest Q values as long as there are enough resources. However, selecting segments with largest Q-values may limit the models ability to explore the solution space by choosing the same segments all the time. Thus, we utilize exploration together with exploitation to ensure that we can explore new solutions while taking advantage of discovered solutions. 

% The value of $\gamma$ in Q-Learning indicates the probability of taking a random action (i.e., selecting a random segment to entangle). We set the initial value of $\gamma = 0.5$ and gradually lower over time. This exploration technique strikes a balance between taking advantage of discovered (optimal) actions and searching for more (optimal) actions.

% Successful attempts change the network state to $S^{(t +1)}$.
% Next, we calculate the reward similar to Q-learning at the end of each time slot. The pair of $(S^{(t)} , S^{(t +1)} , reward , a_m^{(t)})$ is kept in a buffer to train the model in batches.

% Figure \ref{fig:preswap_1} illustrates an example of selected links with deep Q-learning. The model predicted to create entanglement between A-D and C-F, so we conducted swapping operations (i.e., BSM) between links A-C and C-D and links C-D and D-F. Virtual links are created after successful swapping operations.
% \begin{figure}[t]
% \begin{center}
% %\vspace{-2mm}
% \includegraphics[keepaspectratio=true,angle=0,width=0.6\linewidth] {figures/preswap_1.png}
% \vspace{-3mm}
% \caption{Virtual links A-D and C-F are created after swapping entanglement between links A-C, C-D and C-D, D-F. Doing so reduced the average distance between nodes, which would increase the likelihood of success for future request.}
% \label{fig:preswap_1}
% \vspace{-6mm}
% \end{center}
% \end{figure} 

\section{Evaluations}
\begin{figure}[t]
\begin{center}
%\vspace{-2mm}
\includegraphics[keepaspectratio=true,angle=0,width=0.7\linewidth] {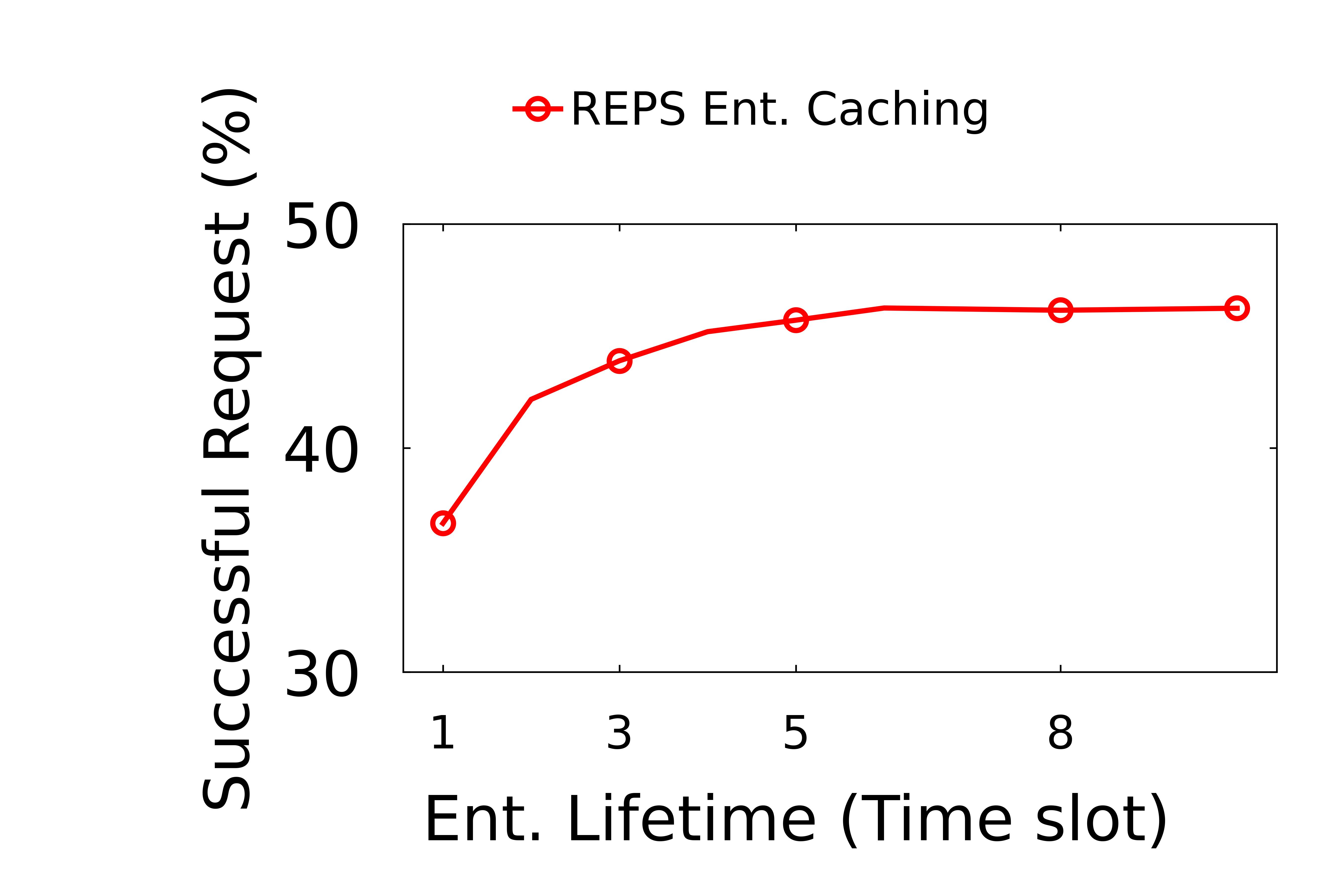}
\vspace{-3mm}
\caption{Impact of entanglement lifetime on the performance of \name.}
\label{fig:lifetime_success}
\vspace{-5mm}
\end{center}
\end{figure}

% \begin{figure*}
% \begin{center}
% \subfigure[Number of requests]{
% \includegraphics[keepaspectratio=true,angle=0,width=.31\linewidth] {figures/seer_req_success.png}
% \label{fig:seer_req_success}}
% \hspace{-6mm}
% \subfigure[Swap probability]{
% \includegraphics[keepaspectratio=true,angle=0,width=.31\linewidth] {figures/seer_q_success.png}
% \label{fig:seer_q_success}}
% \hspace{-6mm}
% \subfigure[Entanglement probability ]{
% \includegraphics[keepaspectratio=true,angle=0,width=.31\linewidth] {figures/seer_alpha_success.png}
% \label{fig:seer_alpha_success}}
% \caption{Performance analysis of entanglement caching and proactive entanglement generation on SEER quantum routing algorithm.}
% \vspace{-3mm}
% \label{fig:seer_result}
% \end{center}
% \end{figure*} 

% \begin{figure*}
% \begin{center}
% \subfigure[Number of requests]{
% \includegraphics[keepaspectratio=true,angle=0,width=.31\linewidth] {figures/reps_req_success.png}
% \label{fig:reps_req_success}}
% \hspace{-6mm}
% \subfigure[Swap probability]{
% \includegraphics[keepaspectratio=true,angle=0,width=.31\linewidth] {figures/reps_q_success.png}
% \label{fig:reps_q_success}}
% \hspace{-6mm}
% \subfigure[Entanglement probability]{
% \includegraphics[keepaspectratio=true,angle=0,width=.31\linewidth] {figures/reps_alpha_success.png}
% \label{fig:reps_alpha_success}}
% \vspace{-2mm}
% \caption{Performance analysis of entanglement caching and proactive entanglement generation on REPS quantum routing algorithm.}
% \vspace{-6mm}
% \label{fig:reps_result}
% \end{center}
% \end{figure*} 

\begin{figure*}
\begin{center}
\subfigure[Number of requests]{
\includegraphics[keepaspectratio=true,angle=0,width=.31\linewidth] {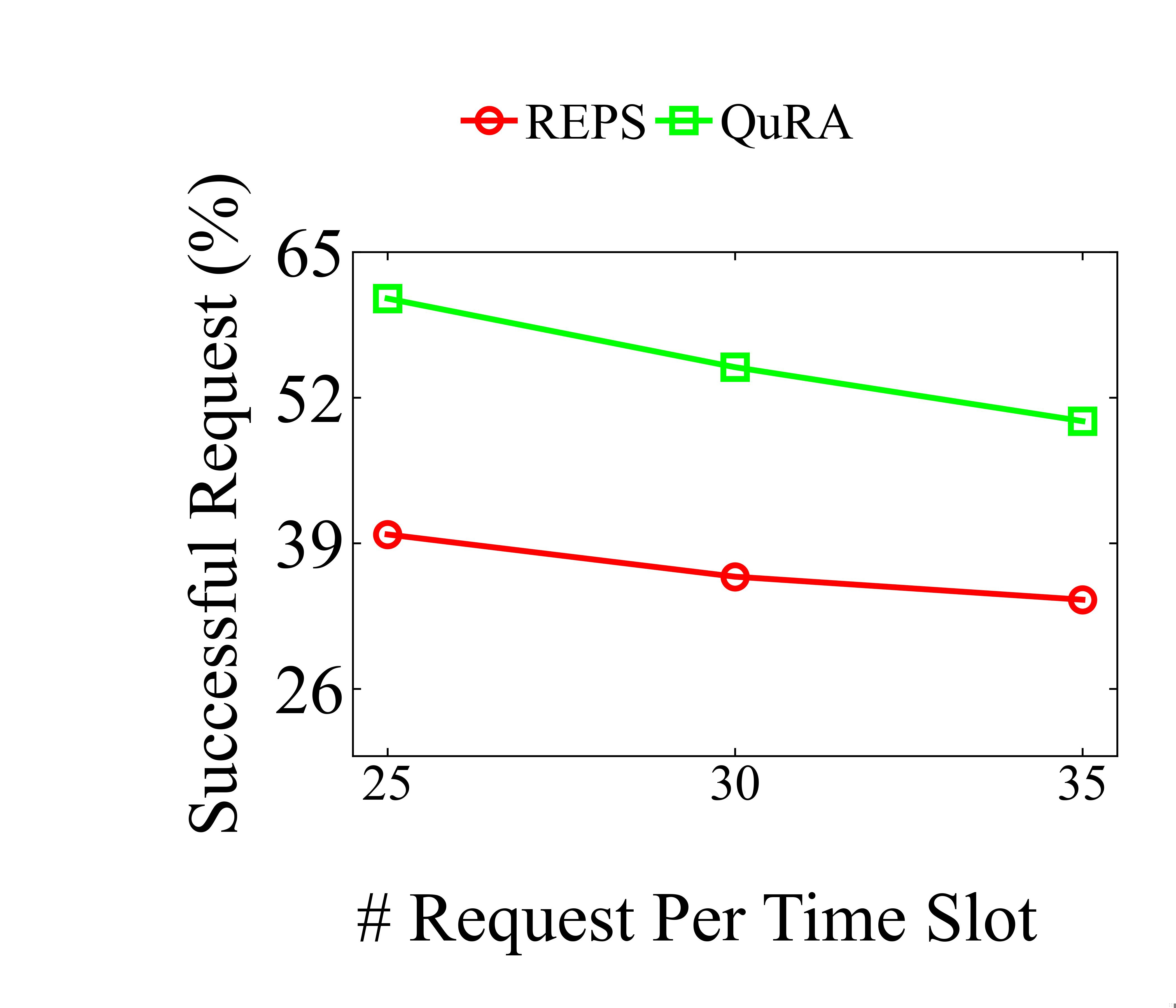}
\label{fig:reps_req_success}}
\hspace{-6mm}
\subfigure[Swap probability]{
\includegraphics[keepaspectratio=true,angle=0,width=.31\linewidth] {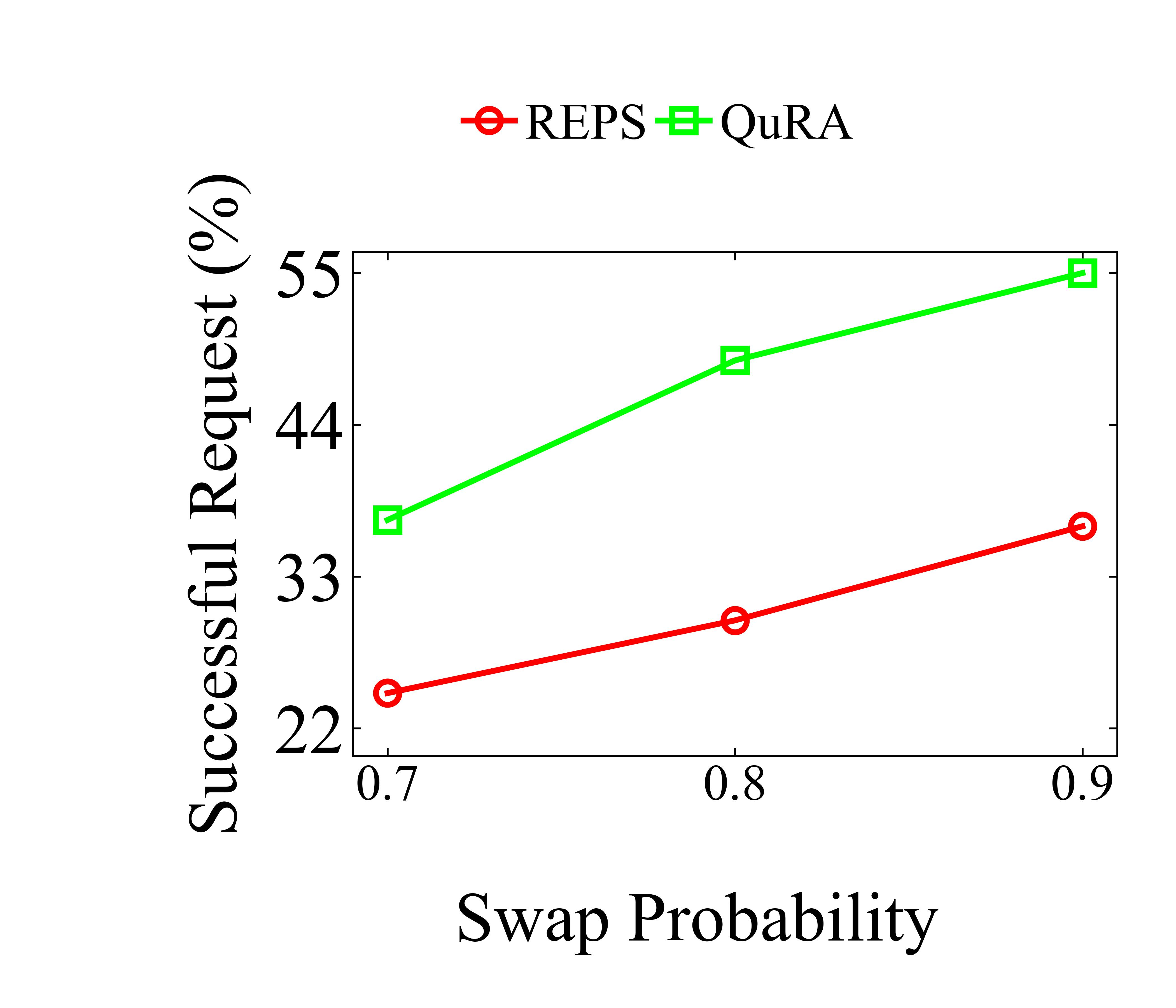}
\label{fig:reps_q_success}}
\hspace{-6mm}
\subfigure[Entanglement probability]{
\includegraphics[keepaspectratio=true,angle=0,width=.31\linewidth] {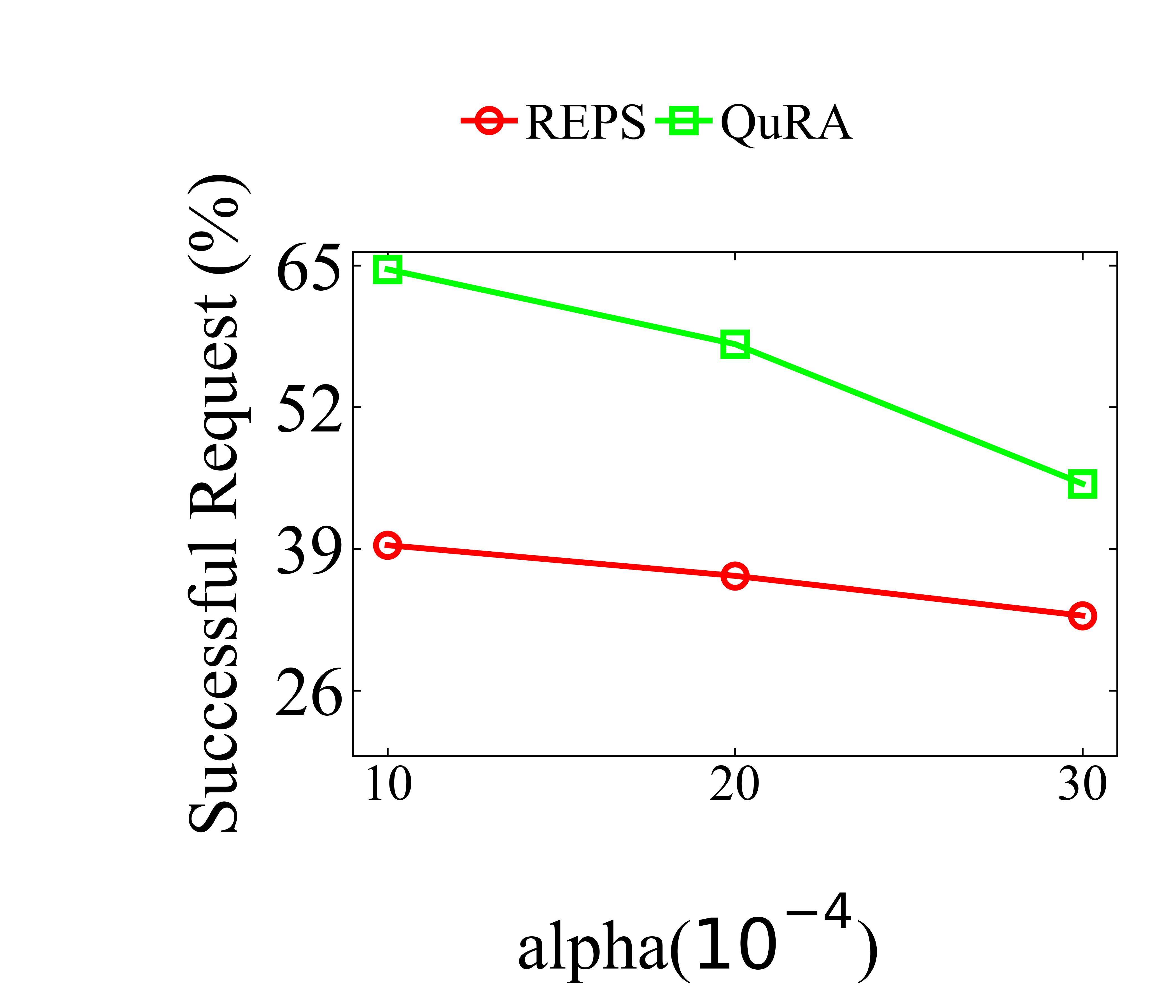}
\label{fig:reps_alpha_success}}
\vspace{-2mm}
\caption{Performance comparison of REPS and \name for varying request count, swap probability, and entanglement probability values.}
\vspace{-6mm}
\label{fig:reps_result}
\end{center}
\end{figure*} 

% We integrated the proposed entanglements proactive entanglement generation with caching and swapping methods into two quantum routing algorithms algorithms, namely REPS \cite{zhao2021reps} and SEER \cite{huang2022seer}. REPS uses linear programming to identify optimal links for each request in a given time slot. After entanglement is attempted on selected links, it tries entanglement swapping using successfully entangled links to create end-to-end entanglement for each request. On the other hand, SEER exploits the social relationship of the nodes and divides the requests (if possible) into two parts. Then, it tries to generate entanglement for each part separately. It then attempts entanglement swapping to combine the both parts. As a result, it takes several time slots to satisfy a given connection request. To better fit and observe the effects of our third approach, link selection with reinforcement learning, we apply this approach only on REPS because in REPS because link selection for entanglement and swapping based on the successful entangled links are exclusive in REPS but in SEER shortest paths are defined first and selection of links for entanglement is based on the shortest paths and decision for swapping also depends on the shortest path selection.

To evaluate the performance of RL-based link selection and proactive entanglement swapping solutions, we integrated \name into quantum routing algorithm, REPS \cite{zhao2021reps}. REPS uses two separate ILPs for link selection and path selection operations. We replaced the first ILP in REPS with \name's RL based link selection algorithm. Then, we incorporated proactive entanglement swapping solution into REPS's path selection algorithm. As a result, proactive entanglement generation receives the output of RL-based link selection algorithm (i.e., the network with a set of successfully entangled links) as input and produces output (a new network with a combination of entangled segments and links). Its output is then consumed by the second ILP of REPS algorithm to choose the optimal paths for given connections.

\subsection{Evaluation Methodology}

To construct the network, we employed the Waxman model and adhered to the simulation methodology outlined in \cite{shi2020qcastpass,zhao2021reps,huang2022seer}. In accordance with these simulation parameters, we placed 50 quantum nodes within a rectangular area measuring 2000 km x 4000 km. Each link has a randomly generated qubit transmission capacity between 3-7 and each node has randomly generated quantum memory capacity between 10-14. The probability for entanglement generation was formulated as $\kern-0.15em P(u,v) = e^{-\alpha \cdot l(u,v)}$, where $l(u,v)$ represents the Euclidean distance between two nodes. A typical value for the parameter $\alpha$ was set to $0.002$. With $\alpha = 0.002$ and $l(u,v)=100km$ the value of $\kern-0.15em P(u,v) = e^{-\alpha \cdot l(u,v)}$ is $0.819$.  The success probability for entanglement swapping was configured to $0.9$. Our simulations, on average, ran for 200,000 time slots, and the results are averaged over $10$ trials for each outcome. In Q-value updating phase of Q-learning, we use learning rate of $\beta = 0.1$ and discount factor of $\gamma = 0.95$.
% The simulations were executed on CloudLab \cite{Duplyakin+:ATC19cloudlab}.

% First, we collected results for original REPS algorithm. Then, we compare it against simple entanglement caching solution (REPS-EC) that only stores unused entanglements for up to $10$ time slots. We next compare them against proactive entanglement swapping (PES) with deep Q-learning model that uses original ILP based routing (REPS-PES). We finally replace ILP with Deep RL based optimizer, then again collect results for REPS with and without any caching optimization (REPS-DRL, REPS-DRL-EC, REPS-DRL-PES). 

%First, we collected results for original REPS algorithm and then, we compare it against simple entanglement caching solution (REPS-EC) that only stores unused entanglements for up to $10$ time slots. Then, we compare REPS against our QuRA algorithm which includes RL based Link selection, entanglement caching and proactive entanglement swapping. Next, we compare our RL based link selection algorithm (QuRA-LS) that only replaces ILP with Deep RL based optimizer against entanglement caching solution (QuRA-EC). We finally compare them against proactive entanglement swapping with deep Q-learning model (QuRA-EC).

\subsection{Evaluation Results}
\subsubsection{Entanglement Lifetime}
The impact of entanglement caching strategy is dependent on the duration of entanglement. If an entanglement can be sustained for only one time slot, no entangled link can be retained in the cache. However, if entanglement can persist longer, then caching can enhance the performance. We illustrate how entanglement duration impacts the performance of quantum routing in Figure \ref{fig:lifetime_success}. When entanglement is sustained for two time slots, the performance improves by 7\%. The improvement rate increases to 19.21\% when the entanglement lifetime is set to six or more time slots. It is evident, entanglement caching leads to much better performance, and the success rate plateaus at  six time slots, indicating that all cached entangled links are used within six time slots. Given that the entanglement lifetime can last for $10$ seconds \cite{bradley2019entanglement10sec} and most quantum routing algorithms consider each time slot as one (or less) second, we set the default entanglement lifetime as $10$ time slots.

\subsubsection{Request Count} 
% Figures \ref{fig:seer_req_success} and \ref{fig:reps_req_success} illustrate the impact of  the number of new requests on SEER and REPS algorithms, respectively. As the number of requests increase, we observe  an increase in the performance for both algorithms. While the original SEER implementation attains less than $50\%$, Deep Q Learning based Proactive Entanglement Swapping (PES-DeepQRL) can succeed more than 70\% of requests. Caching unused entanglements and proactively entangling links (i.e., no swapping)(Ent. Caching) improves the performance of SEER by 22.34\%. the performance improvement reached to 41.79\% when entanglement caching is complemented with proactive entanglement swapping via Q-learning (PES-QRL).

% For REPS, the performance improved by around 18.34\% with entanglement caching alone (Figure~\ref{fig:reps_req_success}). Q-learning based proactive entanglement swapping increases the request success rate by 52.55\%. Unlike SEER which finds only one path for each connection (one for source to intermediate and one for intermediate to destination), REPS identifies multiple paths, allowing proactive entanglement swapping to attain better performance. 

Figure \ref{fig:reps_req_success} illustrates the impact of  the number of new requests. As the number of requests increase, we observe a decrease in the performance in terms of serving the requests. \name outperforms REPS in term of serving requests by 52.55\%. This can be attributed to \name's entanglement caching and proactive swapping solutions as it helps us to reduce the number of links to entangle and swap at the time of request.

\subsubsection{Swap Probability}
% The success of quantum entanglement routing algorithms is significantly influenced by the probability of successfully swapping entangled links. As depicted in Figure \ref{fig:seer_q_success}, an increase in the swap probability correlates with improved performance for SEER algorithm. Proactive entanglement leads up to to 55\% improvement over the original SEER algorithm. Similarly, in Figure \ref{fig:reps_q_success}, we observe that higher swapping probabilities leads up to 54\% improvement in REPS algorithm. We observe 18\% improvement with  entanglement caching alone. These findings emphasize the notable advantages of our proactive entanglement swapping strategy, particularly in scenarios with higher swap probabilities.

The success of quantum entanglement routing algorithms is significantly influenced by the probability of successfully swapping entangled links. As depicted in Figure \ref{fig:reps_q_success}, an increase in the swap probability correlates with improved performance for REPS algorithm. \name leads up to to 54\% improvement over the original REPS algorithm. As \name includes proactive entanglement swapping and entanglement caching on top of RL based link selection for entanglement, this finding emphasizes the impact of our proactive entanglement swapping strategy, particularly in scenarios with higher swap probabilities.

\subsubsection{Entanglement Generation Probability}

The probability of successful entanglement generation affects the performance of quantum routing algorithm. Figure \ref{fig:reps_alpha_success} shows that as the success rate of entanglement reduces (entanglement success probability has inverse relation to $\alpha$),  the rate of successful requests also decreases. \name improves the performance over REPS by up to 52.38\%. The performance difference is higher when entanglement generation probability is higher since entanglement caching and proactive swapping are able to cache more entangled links and create more entangled segments.

\begin{figure}
\begin{center}
%\vspace{-2mm}
\includegraphics[keepaspectratio=true,angle=0,width=0.65\linewidth] {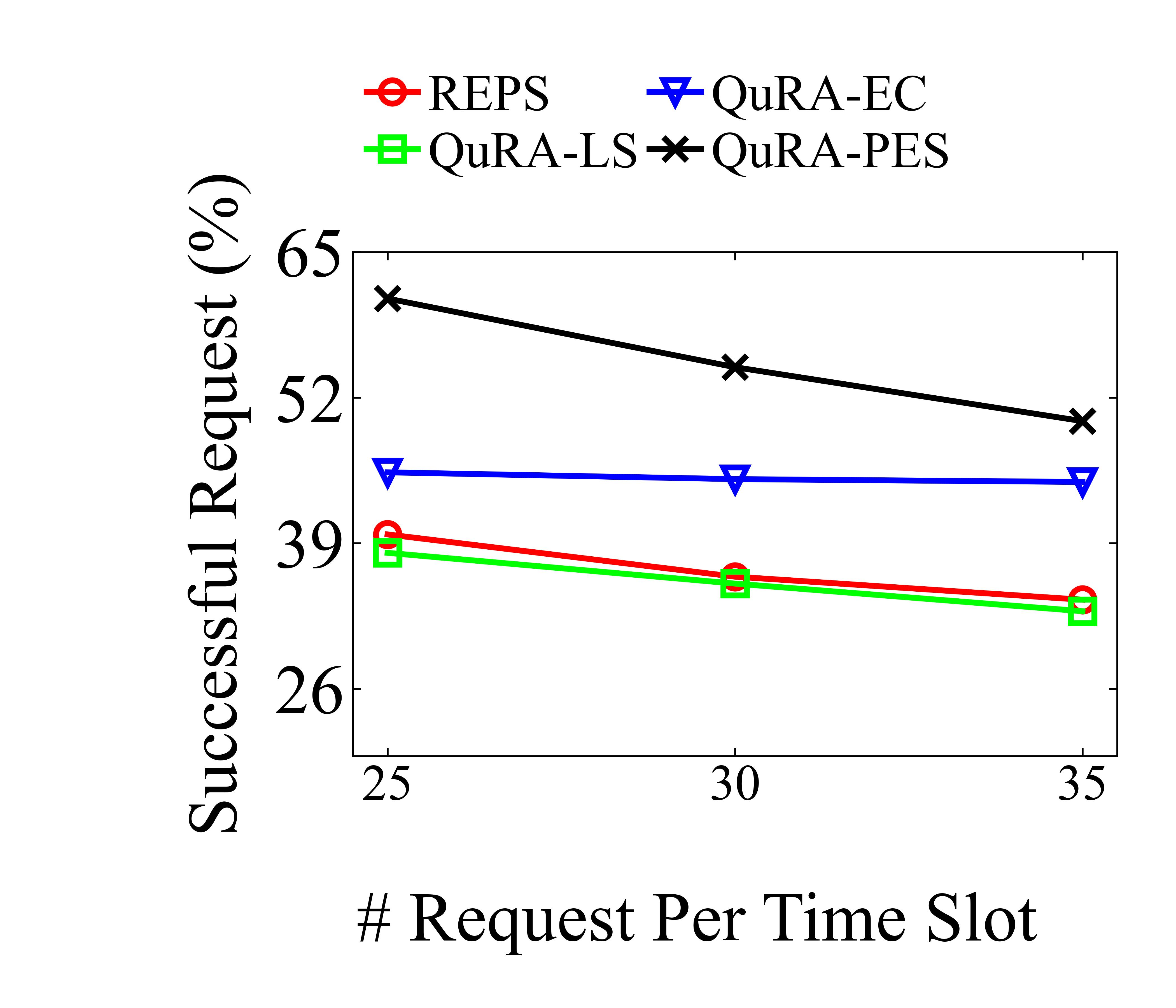}
\vspace{-3mm}
\caption{Performance impact of link selection, entanglement caching and proactive entanglement swapping in \name. }
\label{fig:QuRA_req_success}
\vspace{-6mm}
\end{center}
\end{figure}

\subsection{Ablation Study} 

 Figure \ref{fig:QuRA_req_success} shows the impact of entanglement caching and proactive entanglement swapping for \name. We observe that RL-based link selection for entanglement generation yields similar performance to REPS. Entanglement caching with RL based link selection improves the performance by 18.34\%. We get the best performance combining RL based link selection with caching and proactive swapping. While entanglement caching improves the performance by 15-20\% over REPS, Deep Q-learning based proactive entanglement swapping increases the request success rate by up to 52.55\%, demonstrating the effectiveness of proactively swapping entanglement over frequently accessed links.
 
\subsubsection{Execution Time} REPS uses ILP to obtain near optimal performance~\cite{zhao2021reps}. In Figure \ref{fig:reps_result}, we show that by replacing ILP with Deep RL, \name attains a similar performance with significantly lower execution time. The reinforcement learning approach reduces the runtime to select links for entanglement as much as 20 times than the ILP approach in REPS. We observe that with increasing request per time slot, the runtime remains almost constant in our solution compared to linear increase with ILP.

\begin{figure}[t]
\begin{center}
%\vspace{-2mm}
\includegraphics[keepaspectratio=true,angle=0,width=0.65\linewidth] {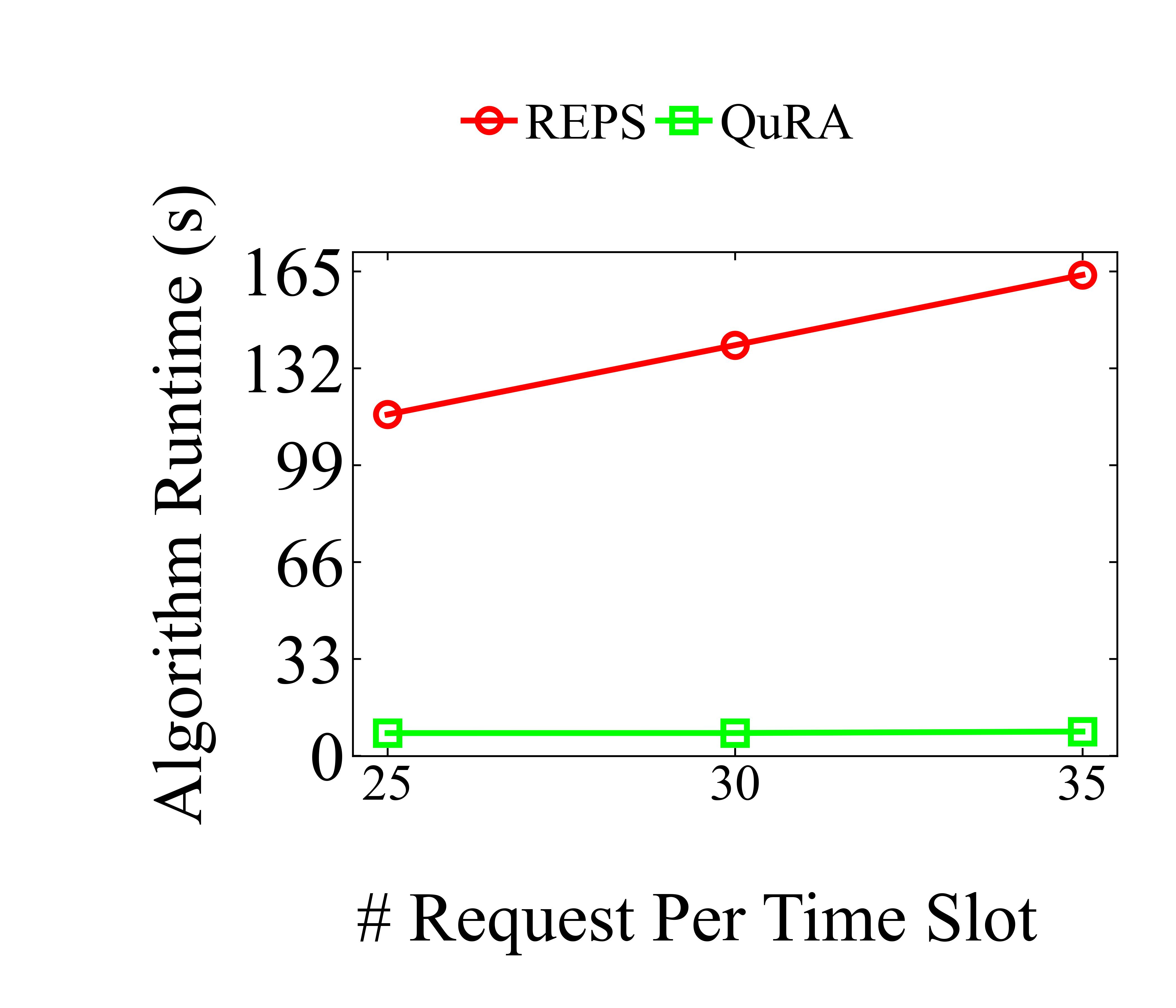}
\vspace{-3mm}
\caption{Execution time for REPS and \name for link selection phase. While REPS relies on Integer Linear Programming, \name takes advantage of Reinforcement Learning to lower the execution time without sacrificing the performance. }
\label{fig:reps_runtime}
\vspace{-6mm}
\end{center}
\end{figure} 

\iffalse

\subfigure[Entanglement lifetime ]{
\includegraphics[keepaspectratio=true,angle=0,width=.24\linewidth] {figures/seer_life_success.png}
\label{fig:seer_life_success}}

\subfigure[Entanglement lifetime ]{
\includegraphics[keepaspectratio=true,angle=0,width=.24\linewidth] {figures/reps_life_success.png}
\label{fig:reps_life_success}}

\fi

\section{Conclusion}

In this paper, we introduced a fast and scalable link selection mechanism for large quantum networks. We applied reinforcement learning (RL) to select potential links for entanglement generation and swapping. RL reduces the execution time by more than 20$\times$ compared to integer linear programming based solution. Additionally, by taking advantage of entanglement caching and proactive swapping, we can improve the performance of routing algorithms by around $50\%$. The evaluation results shows that RL is able to reduce execution time without compromising the performance, which makes it more practical and feasible for real-world quantum networks.

\bibliographystyle{ACM-Reference-Format}
\bibliography{references}

\end{document}